\documentstyle[12pt]{article}
\renewcommand{\theequation}{\arabic{section}.\arabic{equation}}

\def\x{{\mbox{\boldmath$x$}}}
\def\u{{\mbox{\boldmath$u$}}}
\def\r{{\mbox{\boldmath$r$}}}

\def\p{{\mbox{\boldmath$p$}}}

\def\v{{\mbox{\boldmath$v$}}}

\def\tom{{\tilde\omega}}
\def\bom{{\bar\omega}}
\def\om{{\omega}}

\def\eps{{\epsilon}}
\def\pb{{\bar p }}
\def\rel{{Re_\lambda}}

\def\begineq{\begin{equation}}
\def\endeq{\end{equation}}

\begin{document}
\bibliographystyle{prsty}
\title{Anisotropy and scaling corrections in  turbulence}
\author{Detlef Lohse $^{1*}$ and Axel M{\"u}ller-Groeling $^2$}
\maketitle

\centerline{$^1$ The James Franck Institute, The University of Chicago,}
\centerline{ 5640 South Ellis Avenue, Chicago, IL 60637, USA}

\vspace{.5cm}

\centerline{$^2$ Department of Physics, University of Toronto,}
\centerline{ 60 St.George Street, Toronto, Ontario M5S 1A7, Canada}

\bigskip

\date{}
\maketitle
\bigskip
\bigskip
\bigskip

%\begin{abstract}
Two parametrizations for second order velocity moments, the Batchelor
parametrization for the r-space structure function and a common
parametrization for the energy spectrum, $E(p)\propto p^{-5/3}\exp(-p/p_d)$,
are examined and compared. In particular, we investigate corrections to the
local scaling exponents induced by finite size effects.
The behavior of local r-- and p--space exponents differs
dramatically.
The
Batchelor type parametrization leads to energy pileups in p-space at
the ends of the ISR. These bottleneck effects result in an extended
r-space scaling range, comparable to experimental ones for the same
Taylor-Reynolds number $\rel$. Shear effects are discussed in terms of
(global) apparent scaling correction $\delta\zeta^{app}$ to classical
scaling. The scaling properties of $\delta\zeta^{app}(\rel )$
differ not only among the parametrizations considered, but also among
r-- and p--space for a {\it given} parametrization. The difference can be
traced back to the subtleties of the crossovers in the velocity moments. Our
observations emphasize the need for more experimental information on
crossovers between different subranges.
%\end{abstract}

\vspace{0.5cm}\noindent

%\begin{narrowtext}
%\narrowtext
%-----------------------------------------------------------------------

\newpage
\section{Introduction}
\setcounter{equation}{0}
In the theory of fully developed turbulence, scaling ranges of velocity
moments in r-- and in p--space
are often put into a one--to--one correspondence with each other.
The two scaling exponents associated with the scaling ranges are believed to
be equivalent.
While this view is correct
for an infinite system (with an infinite scaling range) the relation between
r--space and p--space exponents becomes more complicated for finite Reynolds
numbers. It is the aim of this paper to quantitatively examine these finite
size effects, as they might well be essential to interpret
experiments \cite{ans84,pra92,pra94} and numerical simulations
\cite{ker90,she93}.

We begin our short
review of previous work relevant to the present investigation
by defining the {\it infinite scaling range} exponents. In $r$--space, they are
defined via the velocity structure functions
\begineq
D^{(m)} (r) = \langle (\u (\x + \r ) - \u (\x ))^m \rangle \propto
r^{\zeta_m}.
\label{eq11}
\endeq
{}From a theoretical point of view
\cite{kra59,lvo94a,lvo94b}, the $p$--space scaling exponents
corresponding to the
(discrete) Fourier transformation $\u (\p )$ of $\u (\x )$ are more
easily accessible,
\begineq
 \langle |\u (\p ) |^m \rangle \propto
p^{-\zeta_m}.
\label{eq12}
\endeq
Kolmogorov's classical dimensional analysis of the turbulence problem
\cite{kol41} gives of course the same result for both kinds of scaling
exponents, namely $\zeta_m = m/3$.
Since Landau's famous footnote in ref.\ \cite{ll87} it has been a
matter of interest whether there are scaling corrections
$\delta\zeta_m= \zeta_m - m/3$ to the classical result in the limit of
{\it infinite} Reynolds number $Re$
\cite{kol62,nel94,lvo94a,lvo94b}.

Grossmann and Lohse discovered and analyzed {\it
finite} size scaling corrections
in their reduced wave vector set approximations (REWA, see
\cite{gnlo94a,gnlo94b} and refs.\ therein) of the Navier--Stokes equations.
These calculations fill the huge gap in the
Taylor-Reynolds number $\rel$ between experiments with $\rel >
10^4$ \cite{cha78,pra94} and full numerical simulations, which
reach only $\rel \approx 200$ \cite{she93}. Similar to the
highest $\rel$ experiments, REWA \cite{gnlo94b,gro94b} achieves $\rel >
10^4$, thus
resolving four decades in p-space.
Three different ranges were distinguished in p-space: The stirring
subrange (SSR, small $p$) with slight intermittency (i.e., scaling
corrections), the viscous subrange (VSR, large p) with strong
intermittency, and the inertial subrange (ISR, medium p) with hardly
any intermittency \cite{gnlo94a,gro94b}.
The physical origin of the VSR and SSR scaling corrections was
extensively discussed in \cite{gnlo94a}. The VSR corrections arise
{}from the competition between turbulent energy transfer downscale and
viscous damping \cite{gnlo92b,fri81}. The SSR scaling corrections are
presumably due to the broken symmetry of the Navier-Stokes dynamics
because of the {\it finite size}  of the system: For small $p$ {\it only
downscale} energy transfer is possible, i.e., the translational
invariance and the self similarity of the turbulent flow
is broken by the boundary conditions.
In addition to the investigation of these local scaling corrections, REWA
also offered the opportunity to study the $\rel$--dependence of {\it global}
corrections to classical scaling. Values of $\rel$ ranging from
$10^2$ to $10^4$ could be simulated \cite{gnlo94b,gro94} and it was
shown \cite{gro94} that $\delta\zeta_m \propto \rel^{-3/5}$ due to the
spectral corrections
to classical scaling,
\begineq
\langle | \u ( \p ) |^m \rangle \propto p^{-m/3} \left( 1 + \alpha_m
\left({p\over p_s}\right)^{-2/3} \right) ,
\label{eq13}
\endeq
which result from large scale anisotropy (e.g., shear).
The $p^{-2/3}$ shear correction has first been suggested by Lumley
\cite{lum67}, who employed dimensional analysis,
and was later also found in refs.\cite{kuz81,fal93a,gro94,yak94} with the help
of dimensional analysis in terms of Clebsch variables.
In refs.\ \cite{kuz81,loh94b} the correction has been associated with
conserved helicity flux in p-space.
The parameter
$p_s$ is the typical scale set by the strength of the shear $s$,
$
p_s =\sqrt{s^3/\eps},
$
and $\alpha_m$ a dimensionless parameter, presumably on the order of $1$.
It is not yet clear, whether the second term in eq.(\ref{eq13}) is only a
small--$p$ {\it correction} or whether pure shear energy spectra
$E(p) \propto \langle |\u (\p )|^2\rangle/p
 \propto p^{-7/3}$ exist, i.e., whether $p_s\gg p_L$ can be achieved. Here,
$p_L \equiv 1/L$ is the momentum scale set by the external stirring force.
If pure shear spectra exist, they will probably be more pronounced in cross
spectra \cite{my75} $E_{12}(p)$.
Yakhot \cite{yak94} recently discussed experimental indications for pure
shear spectra. Whether Maloy and
Goldburg's \cite{mal93} scaling of the velocity structure function
$D^{(2)}(r) \propto r^{4/3} $ in Taylor--Couette flow with oscillatory
inner cylinder corresponds to (\ref{eq13}) is also not clear.

A systematic analysis of the properties of $r$-- and $p$--space scaling
exponents in finite--$\rel$ turbulence has been performed by the
present authors
in ref.\cite{amg94a}. We demonstrated that the $r$--space crossover from
the ISR to the VSR and from the ISR to a large--$r$ saturation range can lead
to energy--pileups at both ends of the p-space ISR, the so--called bottleneck
phenomenon \cite{fal94}. In other words, monotonous local $r$--space scaling
exponents may give rise to non--monotonous local $p$--space exponents. Both
physical (based on the conserved energy current in p--space) and formal
explanations for this effect as well as a comparison with available
experimental and numerical data were given in ref. \cite{amg94a}.

In the present paper, we continue and extend our  investigation of
finite size effects on local scaling exponents in $r$-- and
in $p$--space. In the absence of analytical techniques,
that would enable us to
study these questions directly from a dynamical point of view
(which would of course be preferable), our strategy is
the following: Throughout the paper we will
compare two common parametrizations (and finite size corrections thereof)
for the scaling behavior of velocity moments, namely the $p$--space
parametrization (\ref{eq21})
discussed by Foias, Manely and Sirovich \cite{foi90}
(henceforth called 'FMS--Parametrization' for simplicity)
and the
Batchelor--parametrization (\ref{eq26})
\cite{bat51,my75}, common in $r$--space.
The FMS--parametrization has already been discussed in earlier work by
Brachet et al. \cite{bra83} and Frisch et al.\
\cite{fri81}.
We will see that the main difference between FMS-- and
Batchelor--parametrization lies in the description of the crossover from the
viscous to the inertial range. In this paper, these parametrizations are
considered to be two examples for possible crossover scenarios and we hope
that further experimental results will allow for an unambiguous discrimination
between them. We
believe that pointing out the importance and the surprising consequences
of the nature of the crossovers is a main novel aspect of our work, since
up to now research was mainly concerned with infinite $Re$ scaling exponents.

The paper
consists of two major parts: The first one deals with local small--$p$
scaling corrections. The bottleneck energy pileup at the infrared end of
the ISR \cite{amg94a} as well as the small $p$ scaling
corrections found in REWA \cite{gnlo94a}
and also by dimensional analysis in terms of Clebsch variables
\cite{gro94,kuz81,yak94} belong to this category. We believe that these are all
manifestations of the broken Navier--Stokes symmetry due to the
boundaries, i.e., large scale anisotropy \cite{gnlo94a}.
To investigate possible relations
among these effects we modify the above--mentioned parametrizations to
describe the scaling corrections from REWA, examine the ensuing consequences
in $r$--space,
%\footnote{While a direct Fourier transformation of the REWA
%velocity modes is possible, the sparseness of large--$p$ wave vectors
%reduces the quality of the scaling properties in $r$--space \cite{gro94b}
%and underestimates small structures.},
and perform a quantitative comparison
for the local scaling exponents $\zeta_2(p)$ resulting from the three different
approaches in refs.\cite{gnlo94a,gro94,amg94a}.

In the second part of the paper we focus our
attention on the Taylor--Reynolds
number dependences of the apparent (global)
scaling corrections $\delta\zeta_2^{app}$
due to shear effects.
While it is probably not too surprising
that Batchelor-- and FMS--parametrization lead to different
$\rel$--dependences of the apparent scaling corrections
$\delta\zeta_2^{app}$, our result, that {\it both}
parametrizations give rise to {\it different} behaviour of
$\delta\zeta_2^{app}$
in momentum and coordinate space, respectively, is certainly unexpected. E.g.,
we find that the $p$--space result $\delta\zeta_2^{app,p}
\propto\rel^{-3/5}$ of
ref.\cite{gro94,lvo94a}
corresponds to $\delta\zeta_2^{app,r}\propto\rel^{-1/2}$ in
$r$--space. This observation can be viewed as yet more evidence for the
fact that finite size effects can render $r$-- and $p$--space exponents
inequivalent.

Specifically, our paper is organized as follows: In Sect.2 we introduce the
FMS-- and Batchelor--parametrizations and review our earlier calculations
\cite{amg94a} comparing both parametrizations in $p$-- and in $r$--space.
Sect.3 contains the corresponding analysis for power
spectra and correlation functions in the time domain \cite{amg94c}.
In Sect.4 we investigate
the consequences of the SSR scaling corrections found in REWA and compare the
small--$p$ scaling corrections of refs.\cite{gnlo94a,gro94,amg94a}. Sect.5 is
devoted to a detailed study of shear effects, in particular, to the apparent
scaling corrections $\delta\zeta_2^{app}$ induced in $r$-- and in $p$--space.
Finally, Sect.6 contains our summary and conclusions.

\section{Batchelor-- versus FMS--parametrization}
\setcounter{equation}{0}
\subsection{Definitions and Fourier transforms}

\noindent
To describe the behavior of energy spectra $E(p)$,
the FMS--parametrization \cite{foi90,fri81,bra83}
\begineq
E_{\hbox{\tiny FMS}}(p) = E_0 \eps^{2/3} p^{-5/3} \exp{(-p/p_d)}
\label{eq21}
\endeq
has frequently been used to interpret experimental \cite{pro91,she93b} and
numerical \cite{she93,gnlo94a,gnlo94b} data.
Here, $E_0$ is the $p$--space Kolmogorov constant, $\eps$ the energy
dissipation
rate, and $p_d$ characterizes the crossover to the viscous range. On the other
hand, measured structure functions $D^{(2)}(r)$ are well described by the
Batchelor parametrization \cite{bat51,my75,ans84,eff87,ben94a,amg94a,sto93},
\begineq
D^{(2)}_{\rm B}(r) = {\eps r^2 / (3\nu ) \over \left( 1 + \left({1\over 3b}
\right)^{3/2} \left( { r\over \eta }\right)^2 \right)^{2/3}},
\label{eq26}
\endeq
where $\nu$ is the viscosity, $\eta=\nu^{3/4}/\eps^{1/4}$ the Kolmogorov
length,
and $b=27 \Gamma (4/3) E_0/5 =
6.0 -
8.4$ the experimentally determined \cite{my75} $r$--space Kolmogorov
constant.
The generalisation of both parametrizations to $\zeta_2 \ne 2/3$ is
straightforward and was considered in \cite{amg94a}.
The essential aspects of our present work do not depend on the precise
value of $\zeta_2$.
Clearly, both parametrizations neglect the finite size of the system
since they do not contain a scale for the external stirring force.
Furthermore, the FMS--parametrization eq.(\ref{eq21}) does not contain any
energy pilup (or bottleneck effect) \cite{fal94,amg94a}. The velocity
structure function for a given energy spectrum can be calculated
through the
Fourier transformation \cite{my75}
\begineq
D^{(2)} (r) = 4\int_0^\infty E(p) \left( 1 - {\sin{(pr)}\over pr}
\right) dp.
\label{eq22}
\endeq
Inverting this equation, i.e., calculating the energy spectrum from a given
structure function, requires a short discussion.
Let us consider turbulence in a large but finite domain, so that
$E(p)\rightarrow 0$ as $p\rightarrow 0$. Then the term
involving no trigonometric function on the rhs of eq.(\ref{eq22})
is finite, $4\int_0^\infty E(p) dp \equiv
D^{(2)}(\infty) < \infty$. Physically, this term corresponds to the total
energy
in the fluid. We can now straightforwardly invert eq.(\ref{eq22}) to give
\begineq
E(p) = -{1\over 2\pi} \int_0^\infty pr \ \sin(pr)
         \left[ D^{(2)}(r) - D^{(2)}(\infty) \right] dr.
\label{eq27b}
\endeq
In the limit of infinite system size $D^{(2)}(\infty)$ grows beyond all bounds,
rendering eq.(\ref{eq27b}) ill--defined at first sight. However, since
$\int_0^\infty pr \ \sin(pr) dr \propto p\delta'(p)$,
this affects only singular
contributions at the origin which we may safely discard. We will therefore
always use eq.(\ref{eq27b}) with the understanding that
$D^{(2)}(\infty) = 0$.
Formally this means nothing more but that the Fourier transformation
of a function will not change (apart from the $\delta$-function) if
the function is shifted by a constant.
With the help of the transformation equations
(\ref{eq22}) and (\ref{eq27b}) we can calculate the structure function
corresponding to the FMS--parametrization (\ref{eq21})
and the energy spectrum associated
with the Batchelor--parametrization (\ref{eq26}), giving
\begineq
D^{(2)}_{\hbox{\tiny FMS}}(r)=  {4 E_0 \Gamma(-2/3) \over r(5/3 )
p_d^{5/3}} \left( {5\over 3} p_dr  -\left(1+p_d^2r^2
\right)^{5/6} \sin{\left( {5\over 3} \arctan (p_dr)
\right)}\right).
\label{eq23}
\endeq
and
\begin{eqnarray}
E_{\rm B}(p) &=& - {1 \over 4\pi} {\eps\over 3\nu} p r_d' {d^3 \over dp^3}
             \int_{-\infty}^\infty { \exp(ipr_d'x) \over (1+x^2)^{2/3}} dx
             \nonumber \\
       &=& E_0\eps^{2/3}p_d^{\prime -5/3} \  A
           \left[ {2\over 3} \tilde{p}^{1/6} K_{11/6}(\tilde{p}) +
                             \tilde{p}^{7/6} K_{5/6}(\tilde{p}) \right],
\label{eq28}
\end{eqnarray}
respectively. Here, we introduced some abbreviations for simplicity:
The dimensionless constant
$A$ has the value $A = (9\Gamma(1/3))/(\sqrt{2\pi} 2^{2/3} 5\Gamma(2/3))$,
$\tilde{p} = pr_d'$, and $K_\nu$ is the
modified Bessel function of third kind \cite{abr70}. The crossover from the
inertial to the viscous range is characterized by $r_d,p_d$ for the
FMS--parametrization and by $r_d',p_d'$ for the Batchelor--parametrization.
The large-- and small--$r$ limits of $D^{(2)}(r)$ are required to give
$D^{(2)}(r)=b(\eps r)^{(2/3)}$ and $D^{(2)}(r)=\eps r^2/(3\nu)$, respectively.
Eq.(\ref{eq26}) is obviously designed to meet these constraints and comparing
the asymptotic relations with the appropriate limits of eq.(\ref{eq23})
fixes $E_0 = 5 b (\Gamma (4/3))^{-1}/27 = 1.74$ and
\begineq
p_d^{-1} = (10 b /27 )^{3/4} \eta \approx 2.34 \eta,
\label{eq25}
\endeq
where we chose $b=8.4$ \cite{my75}.
Now, $r_d$ and $r_d'$ are defined by equating the asymptotic limits,
$\eps r^{2}/ (3\nu) = b (\eps r)^{ 2/3}$, so that we arrive at
\begineq
r_d = r_d^{\prime} = (3b)^{3/4} \eta  \approx 11.25\eta .
\label{eq27}
\endeq
We note, however, that although $r_d$ and $r'_d$ are the same (by
definition),
$D^{(2)}_{\rm B}(r)$
shows a sharper crossover
{}from VSR to ISR than $D_{\hbox{\tiny FMS}}^{(2)}(r)$, i.e.
$D_{\rm B}^{(2)}(r) \ge
D_{\hbox{\tiny FMS}}^{(2)}(r)$
for all $r$.
This can be seen
in Fig.1 of
ref.\cite{amg94a} where we compared
$D^{(2)}_{\rm B}(r)$ with
$D^{(2)}_{\hbox{\tiny FMS}}(r)$. Finally, the $p$--space
crossovers $p_d$ and $p_d'$ are defined by the cutoff
in the exponential decay of the spectrum for large $p$.
Thus, $p_d$ is determined by eq.(\ref{eq21}) and, since
$K_\nu(\tilde{p} =  pr_d') \propto p^{-1/3}
 \exp(-pr_d')$ for large argument, we have
\begineq
p_d^{\prime -1} = r_d' \approx 11.25\eta.
\label{eq29a}
\endeq
Note that the naive expectation that
(p-space crossover) $\times$ (r-space crossover) $\approx 2\pi$,
holds in neither case.
For eqs.\ (\ref{eq21}) and  (\ref{eq23}) we have $r_d p_d\approx 4.8$
whereas for
eqs.\ (\ref{eq26}) and  (\ref{eq28}) we have simply
$r_d^\prime p_d^\prime =1$.

\subsection{Bottleneck phenomenon}

\noindent
In \cite{amg94a} we showed that in contrast to eq.(\ref{eq21}) the
parametrization eq.(\ref{eq26}) contains an important physical phenomenon,
the bottleneck effect \cite{fal94}.
This becomes apparent when comparing the energy spectra
$E_{\mbox{\tiny FMS}}(p)$ and $E_{\rm B}(p)$ in Fig.1.
For small $p$ both functions coincide.
Around $p\approx p_d'$, however,
an energy pileup in the crossover region of $E_{\rm B}(p)$ becomes noticable,
leading to a non--monotonous
logarithmic slope $d\log E_{\rm B}/d\log p$.
The local r- and p-space scaling exponents
\begineq
\zeta_2 (r) = {d\log D^{(2)}(r) \over d\log r}, \quad
-\zeta_2(p)-1 = {d\log E(p) \over d\log p}
\label{eq33}
\endeq
of both the Batchelor-- and the FMS--parametrization are plotted in the
right part of Fig.2.
The minimal local p-space scaling exponent of the spectrum
$E_B(p)$ is 0.44, i.e., the
scaling correction is an order of magnitude larger than the discussed
intermittency corrections \cite{my75,kol62}. This effect
could  explain, that the spectra in numerical simulations
\cite{ker90,she93,her93}
are {\it flatter} than the classical expectation, rather than
being steeper as one might expect from possible intermittency
corrections. To quantify this effect, we have introduced an {\it
effective} (global) p-space scaling exponent $\zeta^{eff} (Re_\lambda
)$ in ref.\ \cite{amg94a} which only slowly approaches its infinite
$Re_\lambda$ value. Very recent measurements of
$\zeta^{eff}(Re_\lambda)$ by Zocchi et al. \cite{zoc94} showed exactly
the same behavior for $\zeta^{eff}(\rel)$ \cite{amg94a}.

Furthermore, in ref.\cite{amg94a} we considered a straightforward
generalization of the Batchelor--parametrization,
\begineq
D^{(2)}_B(r) \propto r^2 \cdot (r_d^{'2} + r^2)^{-2/3}
\cdot (L^2 + r^2 )^{1/3},
\label{batn1}
\endeq
which accounts for the crossover from the ISR to a large--$r$ saturation
range induced by the finite scale $L=1/p_L$ set by the external stirring
force. This second crossover might well not be universal,
but our parametrization agrees reasonably well with available data
\cite{ans84,my75,ben94a}.
Calculating the corresponding spectrum we obtained for $r_d' \ll r$
\begin{eqnarray}
E_B(p)  &=&  {\langle \u^2 \rangle L \over \pi }
\Bigg( - {\Gamma (5/6) \over \Gamma(1/3)} \sqrt{\pi}
\left[
{5\over 9} \pb^2
\phantom{}_1F_2({11\over 6},{5\over 2}, {5\over
  2},{\pb^2\over4})
+
{11\over 405} \pb^4
\phantom{}_1F_2({17\over 6},{7\over 2}, {7\over
  2},{\pb^2\over4}) \right] \nonumber \\
&+& {\pi \over 2}
\left[
{1\over 3} \pb
\phantom{}_1F_2({4\over 3},{2}, {3\over
  2},{\pb^2\over4})
+
{2\over 27} \pb^3
\phantom{}_1F_2({7\over 3},{3}, {5\over
  2},{\pb^2\over4}) \right]
\Bigg)
\label{eqn77}
\end{eqnarray}
where $\pb = p/p_L$ and $\phantom{}_1F_2(a,b,c,z)$
denotes a generalized hypergeometric function
\cite{pru90}.
The most prominent feature of this expression is a second bottleneck pileup
at the infrared end of the p--space ISR.
The local logarithmic slopes of
(\ref{batn1}) and (\ref{eqn77}) are plotted in the left part of Fig.2.
Both bottlenecks have the same physical origin, namely the broken symmetry
due to finite size effects. The symmetry breaking scale is introduced by
the stirring force and the finite size of the vessel, wind channel,
or atmosphere \cite{ans84,pra94}
at the infrared end of the spectrum, and by viscosity at the large--$p$
end of the ISR.
Formally, both bottleneck energy pileups originate from the sharp r--space
crossovers defined by the Batchelor--parametrization eq.(\ref{batn1}).
The physical explanation \cite{gnlo94b,amg94a} builds on the
constant energy flux
$T(p) \sim p u(\p)\int dp_1 dp_2 u(\p_1)u(\p_2)
\delta(\p + \p_1+\p_2)$ downscale
in p-space.
For a detailed discussion of the bottleneck effect we refer
to ref.\cite{amg94a}.

\subsection{Higher order moments}

\noindent
The preceding analysis cannot easily be extended to higher order structure
functions. The connection between, say, the fourth order structure function
$D^{(4)}(r)$ and the corresponding fourth moment of $\u (\p)$,
\begin{eqnarray}
D^{(4)}(r) &\propto& \int\delta (\p_1 +\p_2 +\p_3 + \p_3)
\langle \u (\p_1) \u (\p_2) \u(\p_3) \u(\p_4 ) \rangle \nonumber \\
& & \quad \langle\prod_{j=1}^4 (\exp{(i\p_j\cdot \r)}-1)\rangle_{\rm angle}
d\p_1 d\p_2 d\p_3 d\p_4,
\label{neq1}
\end{eqnarray}
is considerably more complicated than eq.(\ref{eq22}). Therefore we have to
restrict ourselves to a few general remarks.

Neglecting intermittency corrections we assume as a first approximation
that
$\v_r(\x ,t) = \u (\x+\r ,t)-\u(\x ,t)$
and $\u (\p ,t)$ are Gaussian distributed, so that we may
simply factorize
higher moments (for even $m$),
\begin{eqnarray}
\langle |\u (\p )|^m\rangle &\propto&
\langle |\u (\p )|^2\rangle^{m/2} , \nonumber\\
D^{(m)} (r) &\propto& \left( D^{(2)}(r) \right)^{m/2}.
\label{eq212}
\end{eqnarray}
Note, however, that the above assumptions are not independent. In a completely
homogeneous medium the second moments in p--space are local
(i.e., $\langle \u^*(\p )\u (\p ')\rangle \propto \delta(\p - \p ')$) and the
second line in eq.(\ref{eq212}) is a direct consequence of the first one.
Of course, the assumption of Gaussian factorization in eq.(\ref{eq212}) is
at variance with the fact that odd moments do not vanish (e.g.,
$D^{(3)}(r)<0$
for large $r$ due to Kolmogorov's structure equation \cite{my75}).

The similar looking factorizations in eq.(\ref{eq212})
lead to quite different results concerning
the m-dependence of the crossovers
$r_d^{\prime (m)}$
and $p_d^{(m)}$ (or
$p_d^{\prime (m)}$) between VSR and ISR.
These lengths are defined as above
by matching the asymptotic behavior for large and for small $r$
and by the cutoff in the
exponentials, respectively.
In r-space we get
\begineq
r_d^{\prime (m)} =
r_d^{\prime (2)} =
r_d^{\prime } = \hbox{constant}
\label{eq214davor}
\endeq
for all $m$, which is in agreement with
recent measurements \cite{sto93}, while
in p-space
\begineq
p_d^{ (m)} = 2 p_d^{ (2)} / m = 2p_d / m
\label{eq214}
\endeq
becomes smaller with increasing $m$.
(The same relation holds for $p_d'$).
Eq.\ (\ref{eq214}) has been numerically confirmed to a high precision
\cite{gnlo94a}. Thus, for increasing $m$ the ISR becomes smaller and
smaller in p-space, whereas it remains invariant in r-space.
Technically, this is due to the fact that we compare two power laws in r--space
and a power law with an {\it exponential} in p--space. An intuitive
understanding,
we hope, is provided by the following remark:
Raising
$D^{(2)}(r)$ to some power smoothes the transition from VSR to ISR and
consequently
reduces the corresponding spectral strength at large values of $p$.

\section{Frequency spectra}
\setcounter{equation}{0}
Instead of performing our analysis for the
structure function $D^{(2)}(r)=D(r)$
(for simplicity, we drop the index 2)
and the spectrum $E(p)$ we can do the same for the
{\it longitudinal} structure function
$D^{(2)}_L(r)= D_L(r)$  and the {\it longitudinal} spectrum
$E_1(p)$ \cite{my75}, which are connected with the
experimentally most easily
accessible time structure function $D(\tau) = \langle
(u_1(t+\tau) - u_1(t))^2\rangle$ and its frequency power spectrum
\begineq
P(\omega ) = - {1\over \pi} \int_0^\infty d\tau \cos (\omega \tau )
D(\tau )
\label{freq1}
\endeq
by Taylor's
hypothesis \cite{tay38,my75}.

As Batchelor's parametrization is also an excellent fit to the
directly measured time structure function
$D(\tau)$ \cite{my75,ben94a,amg94a},
we want to give the corresponding frequency spectrum $P(\omega)$
for completeness. Again we restrict ourselves to classical scaling.

First, for the VSR-ISR crossover
\begineq
D(\tau) \propto {\tau^2 \over (1+(\tau/\tau_d)^2)^{2/3}},
\label{freq2}
\endeq
we obtain from eq.\ (\ref{freq1})
\begineq
P(\omega ) \propto -\bom^{-5/6} K_{5/6}(\bom) + \bom^{1/6}
K_{11/6}(\bom).
\label{freq3}
\endeq
Here, $\tau_d$ characterizes the VSR-ISR crossover and
$\bom=\omega \tau_d$.
The local  logarithmic slopes $\zeta(\tau)$ and $\zeta(\om)$
of (\ref{freq2}) and (\ref{freq3})
(defined as in (\ref{eq33})) are plotted in the
right part of Fig.3. Similar to $\zeta(p)$, the local slope $\zeta(\om)$ is
again
non--monotonous, reflecting the bottleneck phenomenon.
But for $P(\omega)$ (and thus $E_1(p)$) it is only half as
large as above:
The local exponent $\zeta(\om)$ has a minimum of
0.56, compared to 0.44 of $\zeta(p)$, found in
\cite{amg94a}, see also Fig.\ 2.
%This demonstrates (more examples will follow),
%how crucial the p-space crossover behavior (here the strenght of the
%bottleneck
%phenomenon) depends on subtleties of the r-space crossover (here,
%whether $D(r)$ or $D_L(r)$ is better described by the Batchelor
%parametrization).
% diese aussage laesst sich nur machen, wenn man E(k) aus E_1(k) berechnet.

Next, for the ISR-large $\tau$ saturation range we assume
(again led by experimental data \cite{ans84,eff87,my75}),
\begineq
D(\tau) \propto \tau^{2/3} \left( 1+(\tau/\tau_L)^2\right)^{-1/3}.
\label{freq4}
\endeq
This second
crossover is governed by the large eddy turnover time--scale $\tau_L$.
The corresponding spectrum reads
\begineq
P(\omega ) \propto
{\Gamma(5/6)\over \sqrt{\pi}\Gamma(1/3)} \left(
\phantom{}_1F_2({5\over 6},{3\over 2}, {3\over  2},{\tom^2})
+{20\over 27} \tom^2 \
\phantom{}_1F_2({11\over 6},{5\over 2}, {5\over  2},{\tom^2})
\right) -
{1\over 3} \tom \
\phantom{}_1F_2({4\over 3},{2}, {3\over  2},{\tom^2}),
\label{freq5}
\endeq
where $\tom = \omega \tau_L/2$.
The local logarithmic
slopes of (\ref{freq4}) and (\ref{freq5}) are plotted in the
left part of Fig.3. The maximal value of $\zeta (\omega)$ is 0.73
instead of 0.77 found for $\zeta(p)$
in \cite{amg94a}. Again, the bottleneck effect
is smaller in $\omega$-space than in p-space.

Our Fig.3 has to be compared with the experimental local slope
of $P(\omega )$, which
we show in Fig.4, taken from Praskovsky and Oncley's recent paper
\cite{pra94}. While the comparison is definitely not conclusive due to
the experimental noise, there nevertheless seems to be a certain tendency
towards the formation of energy pilups at both ends of the ISR scaling range.
For a discussion of the quantitative discrepancy we refer to section 6.

\section{Small--$p$ scaling corrections}
\setcounter{equation}{0}

\noindent
In this section we examine the consequences of the $p$--space SSR scaling
corrections found by Grossmann and Lohse \cite{gnlo94a,gnlo94b}
in their numerical
analysis. To this end we modify the energy spectra $E_{\rm B}(p)$ and
$E_{\mbox{\tiny FMS}}(p)$ calculated in section 2 to
include these corrections
at the infrared end of the ISR. Two major problems arise in the course of this
procedure: First, we have to introduce the external stirring force scale into
the $p$--space parametrizations to define a finite range
for the scaling corrections. Second, we have to calculate the structure
functions corresponding to the modified, more complicated energy spectra.
In the following we explain how we deal with these problems, present our
results, and compare our findings with other small--$p$ corrections
\cite{gro94,amg94a} that are being discussed.

\subsection{Infrared cutoff}

\noindent
Apart from our discussion of the small--$p$ bottleneck
we have considered only ideal turbulence in an infinite spatial domain up to
now. Real turbulence is restricted to a finite range, i.e., there is a maximal
length scale $L$ or, equivalently, a small wave vector cutoff
$p_L\equiv
1/L$. In
order to discuss the small--$p$ bottleneck in section 2 the Batchelor
parametrization was generalized to include this length scale. Here, we have to
modify the parametrizations  $E_{\rm B}(p)$ and
$E_{\mbox{\tiny FMS}}(p)$ of the energy spectra accordingly.
Therefore we multiply
the spectra by $(2/\pi) \arctan((p/p_L)^{11/3})$.
This amounts to imposing energy equipartition
$E(p) \propto p^2$ on
the unforced wave vector modes \cite{my75,ors77,nel94}
with $p\ll p_L$.
Note, however, that the small--p behavior of the spectrum in Navier--Stokes
dynamics has not been firmly established up to now \cite{nel94} and we adopt
energy equipartition for small $p$
only as one of several possible scenarios. We will come
back to this point at the end of section 4.2.
As an immediate consequence of the cutoff the
corresponding structure functions saturate for large $r$ at the constant
value $D^{(2)}(\infty) \approx D^{(2)}(L)$.

The finite maximal length scale $L$ allows for the introduction of the Reynolds
number $Re$ or alternatively the Taylor-Reynolds number $Re_\lambda = \lambda
u_{1,rms}/\nu$, where $\lambda = u_{1,rms } / (\partial_1 u_1)_{rms}$
is the Taylor length
and $\nu$ the viscosity. Let us express $p_d$ and $p_d^\prime$
in terms of $L$ and $Re_\lambda$. We have $\eps = c_{\eps}
u^3_{1,rms}/L$ with $c_\eps\approx 1$ known from grid turbulence
experiments \cite{sre84}. Here, we neglect the $Re_\lambda$-dependence
of $c_\eps$ for small $Re_\lambda$ \cite{loh94a}.
Note, that $c_\eps$ can directly be connected with the Kolmogorov
constant $b$ \cite{loh94a,amg94a,gro94c}.
On the other hand
$\eps = 15\nu (\partial_1u_1)^2_{rms}$ \cite{my75}.
Using these relations we finally get
$\eta = 15^{3/4} c_\eps^{-1} L Re_\lambda^{-3/2}$  or,
with $c_\eps \approx 1$ and the relations
(\ref{eq25}) and (\ref{eq29a})  for $p_d$, $p_d^\prime$,
\begineq
p_d^{-1}\approx 18 L Re_\lambda^{-3/2},
\label{eq31}
\endeq
\begineq
p_d^{\prime -1}\approx 86 L Re_\lambda^{-3/2}.
\label{eq32}
\endeq
This establishes the connection between the length scales $r_d$
(or $r_d'$) and
$L$, and the Taylor--Reynolds number $\rel$.

\subsection{REWA scaling corrections and structure functions}

\noindent
In the reduced wave vector set approximations of the Navier-Stokes
equation, Grossmann and Lohse \cite{gnlo94a} found deviations
$\delta\zeta_m (p)$ from classical scaling when {\it locally} fitting
the spectra. These deviations  occur
-- as already discussed in the introduction -- only
for small $p$ (SSR intermittency)
and for large $p$ (VSR intermittency), whereas no
scaling corrections were found in the p-space ISR \cite{gnlo94a}.
To investigate how the p-space scaling corrections act in r-space, we
model a spectrum according to the numerical results in
\cite{gnlo94a} and numerically Fourier transform it into r-space.
We will focus attention on p-space SSR scaling corrections.
A short calculation reveals that for any function $E(p)$ with
$-\zeta(p) -1 =d\log E/d\log p$ a modification defined by
\begineq
\tilde{E}(p) = E(p) \
{ (p_b^\beta + p^\beta)^{\delta/\beta} \over p^{\delta}}
\label{eqamg1}
\endeq
leads to a local exponent of the type
\begineq
\tilde{\zeta}(p) = \zeta(p) + {\delta \over 1+(p/p_b)^\beta}.
\label{eqamg2}
\endeq
This is exactly what we want: Assuming that $\zeta(p)=2/3$ we get the
modified exponent $\tilde{\zeta}(p) = 2/3 + \delta$ for small $p$ until
the infrared cutoff sets in.
Furthermore the scheme is very flexible. We can introduce
positive and negative corrections $\delta$ and let them become effective
for $p\ll p_b$ or $p\gg p_b$ depending on the sign of $\beta$.
We mention that this procedure, when applied to the exact small--$r$ result
$D(r) \propto r^2$ in order to extend $D(r)$ to the
ISR, immediately leads to the Batchelor parametrization (for
$\beta=-2$
and $\delta = 4/3$).

For our present purposes  we choose
SSR corrections between $\delta = 0.0 - 0.04$,
in the range of intermittency corrections discussed in the literature
\cite{my75,kol62,ans84}.
For large
$p$ we have $\zeta(p) = 2/3 $, (and for very large $p$ the exponential
cutoff  is supposed to become effective).
The crossover is determined by $p_b$, we choose $p_b= 10 p_L - 15 p_L$, i.e.,
allowing for about one decade of large scale intermittency corrections
in p-space as found in \cite{gnlo94a}. The parameter $\beta$ determines the
smoothness of the transition, we choose $\beta=2$. The two spectra
\begineq
\tilde{E}_{\mbox{\tiny FMS}}(p) = E_{\mbox{\tiny FMS}}(p)
{(p_b^2 + p^2)^{\delta/2}\over p^{\delta} }, \quad
\tilde{E}_{\rm B}(p) = E_{\rm B}(p)
{(p_b^2 + p^2)^{\delta/2}\over p^{\delta} }
\label{eqamg3}
\endeq
were each multiplied by the small $p$ cutoff
$(2/\pi)\arctan((p/p_L)^{11/3+\delta})$ and then Fourier
transformed.
The Fourier transformations of $\tilde E_{\rm B}(p)$
and $\tilde E_{\hbox{\tiny FMS}}(p)$
(including the $\arctan$--cutoff)
were performed numerically, employing
a routine designed to cope with the strongly oscillating integrand.
As the strong oscillations in eq.(\ref{eq22}) are exponentially damped by
the asymptotic behavior of $\tilde
E(p)$, there are no serious numerical difficulties.
The local scaling exponents for p- and r-space
(defined as in eq.\ (\ref{eq33})) are shown in Fig.\ 5a
and 5b, respectively, for a Taylor-Reynolds number of $\rel=3000$,
which is
in the range of typical experiments
\cite{ans84}.

Let us first discuss the results for the FMS parametrization. Without
any small $p$ corrections (\ref{eqamg3}) we have about one decade of
more or less
constant $\zeta (p) \approx 2/3$, Fig.\ 5a.
The corresponding structure function
$D_{\mbox{\tiny FMS}}(r)$, however, does not scale, see Fig.\ 5b,
where no
scaling range can be observed from the local exponent $\zeta(r)$.
This demonstrates that the transformation from $p$-- to $r$--space is not
completely local:
A reasonably well--defined scaling range in $p$--space is
mapped to a $r$--space curve
with only very poor scaling properties or, perhaps more appropriately, with
no scaling range at all.
In the REWA calculations of ref.\ \cite{gro94b} a very similar
behavior of $\zeta(r)$ has been found, which is not surprising, as the
spectra are well parametrized by the FMS parametrization
\cite{gnlo94a,gnlo94b,gro94b}. The poor scaling properties of
$D_{FMS}(r)$  put, in our
opinion, a question mark behind the FMS--parametrization,
since experimental data for structure functions for the same $\rel$
\cite{ans84} exhibit much better scaling.

Only by introducing the small $p$ scaling corrections
(\ref{eqamg3}) to $E_{\mbox{\tiny FMS}}(p)$,
i.e., by making the scaling properties
{\it worse} in p-space (Fig.\ 5a),
we get {\it improved} (but still poor)
scaling of the r-space structure
function (Fig.\ 5b).
As examples, we chose $\delta=0.02$, $p_b/p_L =10$, which is about
what was found in the REWA calculations \cite{gnlo94a,gnlo94b}, and
$\delta=0.04$, $p_b/p_L=15$.
Our  result completely agrees with our findings of section
2: {\it Non--monotonous} $p$--space scaling exponents lead to nicer scaling
properties in $r$--space, i.e., the small $p$ scaling correction
(\ref{eqamg3}) can be interpreted as artificially introduced
bottleneck energy pileup on the infrared end of the $p$--space ISR.

The same analysis is performed for the Batchelor parametrization. Now, due
to the large energy pileup at the ultraviolet end of the
$p$--space ISR, the small--$r$ scaling properties
of $D_B(r)$ are improved considerably.
Yet the $p$-space $\arctan$ infrared
energy cutoff still
corrupts $r$--space scaling properties for large $r$.
Again, this effect can be partly compensated
by local infrared $p$--space corrections of type
(\ref{eqamg3}).
With these corrections, $D_B(r)$ shows  better
scaling properties in r-space (Fig.5b) which are now comparable to the
experimentally realized scaling ranges of about $1.5 - 2$ decades for
that Taylor-Reynolds number $\rel = 3000$ \cite{ans84}.

To summarize: {\it The simple $\arctan$ or $\exp$ cutoffs of the
$p$--space ISR
scaling range lead to unrealistically short $r$--space scaling ranges.
Only the energy pileups at both ends of the $p$--space ISR lead to a
realistic scaling range of the structure function}, if
compared with experiment \cite{ans84}.
{\it Batchelor parametrizations of the crossovers} (\ref{batn1},
\ref{eqn77})
{\it include these energy pileups} and give realistic scaling ranges
for given $\rel$.
Our findings also explain, why the r-space scaling found in REWA is
worse than that in the p-spectra \cite{gro94b}, as the latter is quite well
described by the FMS parametrization \cite{gnlo94a}. \footnote{The
bottleneck pileup found in \cite{gnlo94b} is quantitatively smaller than that
following from eq.\ (\ref{eq28}), so that the FMS-parametrization is
still an appropriate fit.}

Coming back to section 4.1, we briefly mention
that it is only the {\it type} of the ISR to large--$r$ saturation
crossover in (\ref{batn1}) which determines the very small $p\to 0$
behavior for $E_B(p)$. For $\beta=2$ we obtain $E_B(p) \propto p$, for
$\beta=4$ it is $E_B(p) \propto p^2$. It is not clear, which behavior is
the more realistic one. For the Euler equations the latter can be
proven \cite{ors77} to be correct
and simply reflects energy equipartition, but for
the Navier-Stokes dynamics the situation might well be different
\cite{nel94}, in particular, as the thermal energy is  orders of
magnitude smaller than the energy of the large scale eddies.

\subsection{Comparison of infrared scaling corrections}

\noindent
As discussed in the introduction, the large scale anisotropy
(boundary, shear) leads to eq.\ (\ref{eq13}) by dimensional analysis.
We wonder how this result compares with the local slope
$\zeta_2(p)$ resulting from
(\ref{eqn77}) and from the REWA calculations \cite{gnlo94a,gnlo94b}.
Of course we can compare only the crossover region, as
in the original work by Lumley \cite{lum67} and also in the derivation
in \cite{kuz81} the second term ($\propto p^{-2/3}$) in eq.(\ref{eq13})
is considered to be small.

The comparison for the local slopes $\zeta_2 (p)$ is given in Fig.\ 6.
Clearly, although all curves show $\delta\zeta_2 (p)=\zeta_2(p)-2/3 > 0$ at the
infrared end of the p-space ISR,
they do not agree quantitatively.
Note however, that no wave vectors smaller than the
forcing scale were included in REWA. The presence of such wave vectors
leads to the infrared bottleneck effect \cite{amg94a}. It would be
interesting to include such wave vectors in full numerical simulations
or in REWA type simulations and to examine, whether a bottleneck
energy pileup  as in (\ref{eqn77}) or in
(\ref{freq5})
will occur. Also note from Fig.6, that the REWA corrections,
which we introduced into the energy spectra by means of the parametrization
chosen in the preceding subsection, could also be described
by eq.\ (\ref{eq13}) with $\alpha_2 \approx 0.1$. This
confirms and justifies the analysis done in ref.\ \cite{gro94}, see
also section 5. The bottleneck corrections are comparable in size with
the corrections due to eq.\ (\ref{eq13}) (with the arctan cutoff) for
$\alpha_2 = 0.5$.

The quantitative discrepancies in Fig.\ 6 should not be too surprising.
The small $p$ spectral shape is far from being universal due to
different boundary conditions and different kinds of stirring. Yet we
believe that the three kinds of discussed small $p$ scaling
corrections $\delta\zeta_2 (p) > 0$ all have the same origin, namely
the broken symmetry of the Navier-Stokes dynamics for small $p$ and
the broken self similarity of the turbulent flow.

\subsection{Higher order moments}

\noindent
In principle,
the analysis of section 4.2.\ can be repeated
for higher order velocity moments.
In the REWA calculation
\cite{gnlo94a} it was found, that for large $\rel$ higher order moments nearly
factorize into second order moments, of course apart from the p-space
SSR and VSR intermittency corrections. One could assume such a factorization
and parametrize the small $p$ scaling corrections by
\begineq
\zeta_m (p) = {m\over 3} +{\delta_m \over 1+ (p/p_b )^\beta},
\label{eq45}
\endeq
with $\delta_m < 0$ for $m>3$.
We have performed this calculation but refrain from discussing the outcome
in detail because nothing essentially new can be learnt. The results simply
confirm the views already developed when dealing with the second moments.

\section{Shear effects}
\setcounter{equation}{0}

\noindent
In this section we investigate the consequences of an extended shear scaling
range. We will modify the Batchelor-- and the FMS--parametrization in such a
way
that a crossover to a shear scaling range occurs at the scale of the stirring
force $p_L=1/L$. We then proceed to calculate the impact of this
modification on scaling exponents in both $r$-- and $p$--space. In particular,
we focus attention on the  {\it apparent}  scaling
correction $\delta\zeta^{app}(\rel)$ \cite{gro94} as a function of the
Taylor--Reynolds number $\rel$. This correction is induced by the crossover
itself and does not depend on the extension of the shear range.

\subsection{Shear parametrizations}

\noindent
The generalized FMS--parametrization (cf. eqs.(\ref{eq13}) and (\ref{eq21}))
\begineq
\langle | \u ( \p ) |^m \rangle \propto p^{-m/3} \left( 1 +\alpha_m
\left({p\over p_L}\right)^{-2/3} \right) \exp{(-p/p_d)}.
\label{eq51}
\endeq
was shown \cite{gro94} to lead to an apparent scaling correction,
defined by
\begineq
\delta\zeta_m^{app, p } = \hbox{min}_p (\zeta_m (p)) - m/3,
\label{eq53}
\endeq
where, as usual,
$\zeta_m (p) = -d\log \langle | \u (\p ) |^m \rangle /
d\log p $. For $p_L \ll p_d$ it was found that
\begineq
\delta\zeta_m^{app, p } = \mbox{sign}(\alpha_m)
{10\over 9} \left( {9mp_L\over 8p_d}\right)^{2/5}
\alpha_m^{3/5} =
c_m Re^{-3/10} = c_m' Re_\lambda^{-3/5},
\label{eq55}
\endeq
i.e., the apparent scaling corrections vanish with increasing $\rel$
with a $-3/5$ power law for all $m$.
Eq.\  (\ref{eq55}) has been numerically confirmed by reduced wave
vector set calculations \cite{gro94}. The dimensionless constants
$c_m$, $c_m'$, and $\alpha_m$ are found to be negative for $m>3$
\cite{gro94}. This means that higher order moments do not factorize
into second order moments in agreement with the
p-space SSR scaling corrections \cite{gnlo94a} found numerically.
Note that {\it if}
one
assumes factorization of higher order moments in the shear range,
shear can {\it not} account for experimentally observed \cite{ans84}
scaling corrections.
The prediction (\ref{eq55}) \cite{gro94} is also in agreement with
very recent experiments which clearly show a {\it decrease} of
$\delta\zeta_m^{app,p}$ for increasing $\rel$ \cite{kat94}.

The local scaling exponent of eq.\ (\ref{eq51}) is shown in Fig.\ 7a
for $\rel=1500$ (which is determined according to eq.\ (\ref{eq31})).
The apparent scaling correction $\delta\zeta_2^{app,p}\approx 0.06$ is
very large.
Note, however, that according to eq.(\ref{eq55}) $\delta\zeta_2^{app,p}$ is
proportional to $\alpha_2^{3/5}$ and will therfore be smaller for
$\alpha_2<1$.
The $\rel$ dependence is displayed in Fig.\ 8. For
large $\rel$ the asymptotic result (\ref{eq55}) is recovered.

We have mentioned in the introduction that the form of the shear correction
in eq.(\ref{eq13}) is based on dimensional analysis only.
One could argue that similar considerations
as in \cite{lum67} can as well be directly applied
to $r$--space expressions. Assuming the transition to the shear range (with
$r$--space exponent $4/3$) to be of the Batchelor type we are led to the ansatz
\begineq
D_B(r) = {\eps\over 3\nu} r^2 {r_d^{\prime 4/3}\over (r_d^{\prime
2}+r^2)^{2/3}}
             {r_s^{-2/3} \over (r_s^2+r^2)^{-1/3}},
\label{eqamg4}
\endeq
with $r_s = p_s^{-1} = \sqrt{\eps / s^3}$. We assume again that shear sets in
at the stirring scale, hence $r_s=L$.
For simplicity, we again dropped the index 2 of the structure function,
as we will restrict ourselves to $D^{(2)}(r)$ from here on.
Eq.(\ref{eqamg4}) should be viewed
as a generalization of the Batchelor--parametrization in complete analogy to
the generalized FMS--parametrization eq.(\ref{eq51}). The local scaling
exponent of $D_B(r)$ is given by
\begineq
\zeta (r) = 2 + {2\over 3} {r^2\over L^2+r^2} -{4\over 3} {r^2\over
r_d^{\prime 2}+r^2},
\label{eq59}
\endeq
and is shown in Fig.7b. The apparent scaling correction is defined
analogously to
eq.(\ref{eq53}),
\begineq
\delta\zeta^{app, r } = \hbox{min}_r (\zeta (r)) - m/3.
\label{eq57}
\endeq
For the same $\rel = 1500$ as above, $\delta\zeta^{app,r} \approx 0.0028$
is now much smaller than the corresponding value for the generalized FMS
parametrization. This reflects the much better scaling properties of
the Batchelor parametrization (compared to FMS, cf.\ fig.\ 5),
which we have extensively discussed in sections 2 and 4. The $\rel$
dependence of $\delta\zeta^{app,r}$ is  displayed in Fig.\ 8.
For large $L\gg r_d'$ (i.e., for large $\rel$) we obtain
\begineq
\delta\zeta^{app,r} = \sqrt{2} {4\over 3} {r_d'\over L} \propto
Re^{-3/4} \propto \rel^{-3/2}.
\label{eq59s}
\endeq
Hence, also
the asymptotic $\rel$ dependence of $\delta\zeta^{app,r}$ for the
Batchelor parametrization is quite different from that of
 $\delta\zeta^{app,p}$
for the FMS parametrization.
One  might argue that this is not very astonishing
since we compare two different parametrizations. Let us therefore Fourier
transform both parametrizations and reexamine their scaling properties
thereafter.

\subsection{Fourier transforms}

\noindent
Transforming the FMS--parametrization eq.(\ref{eq51}) with the help of
eq.(\ref{eq22}) to r-space leads to
\begin{eqnarray}
D_{\hbox{\tiny FMS}}(r) &\propto&
  { \Gamma(-2/3 ) \over (5/3 ) r
p_d^{5/3}} \left( {5\over 3}p_dr -\left(1+p_d^2r^2
\right)^{5/6} \sin{\left( {5\over 3} \arctan (p_dr)
\right)}\right) \nonumber \\
&+& { \alpha_2\Gamma(-4/3 ) p_L^{2/3} \over (7/3 ) r
p_d^{7/3}} \left( {7\over 3}p_dr -\left(1+p_d^2r^2
\right)^{7/6} \sin{\left({7\over 3} \arctan (p_dr)
\right)}\right)
\label{eq56}.
\end{eqnarray}
The local slope of eq.(\ref{eq56})
for $\rel=1500$, $\alpha_2 =1$, is also plotted in Fig.\ 7a, to compare it
with the slope of
eq.\ (\ref{eq51}).
Now  $\delta\zeta^{app,r}\approx 0.11$ is even larger than
$\delta\zeta^{app,p}$ (both for FMS), which is clearly understandable
{} from Fig.\ 5 because of the even worse scaling properties of the FMS
parametrization in r-space (compared to p-space).

What is more surprising is, that now for the {\it same} (FMS)
parametrization the $\rel$ dependence of
$\delta\zeta^{app,r}$, cf.\ Fig.\ 8, is {\it different}
from that of $\delta\zeta^{app,p}$.
For the former we obtain the asymptotic result
\begineq
\delta\zeta^{app, r } (Re_\lambda)
\propto \rel^{-1/2},
\label{eq58}
\endeq
which is a considerable flatter dependence than (\ref{eq55}).

Finally, we calculate the spectrum corresponding to the
generalized Batchelor parametrization (\ref{eqamg4}).
For $r \gg r_d'$ we can derive an analytical result, which we give in
appendix A for completeness. In the general case, we have to restrict ourselves
to a numerical treatment.
The numerical Fourier transformation (\ref{eq27b}) of (\ref{eqamg4})
(or of (\ref{eq26})) is more delicate than the inverse
transformation (\ref{eq22}) due to the absence of an exponential cutoff.
In appendix B we explain how we convert the strongly oscillating integral to
a rapidly converging one by means of contour integration techniques
\cite{wat22}

The result for the local p-space slope of the generalized Batchelor
parametrization (\ref{eqamg4}) is shown in Fig.\ 7b. Of course it
shows the ultraviolet bottleneck energy pileup \cite{fal94}, which we had
discussed in detail in ref.\ \cite{amg94a}. But now, in addition,
the spectrum shows {\it reduced} spectral strength
at the infrared end of the p-space ISR, i.e., a decreased local slope
$\zeta(p) <2/3$. This effect can be interpreted as, so to say, an {\it
  inverse} bottleneck effect and can both formally and physically be
interpreted along the same line of arguments as the bottleneck
pile{\it ups} discussed above and in
\cite{amg94a}. Formally it reflects the sharp crossover
from $r^{2/3}$ to $r^{4/3}$ scaling in the structure function.
Physically \cite{gnlo94b,amg94a}, the constant energy
flux
$T(p) \sim p u(\p)\int dp_1 dp_2 u(\p_1)u(\p_2)
\delta(\p + \p_1+\p_2)$
downscale
now requires {\it reduced} spectral strength at the infrared end of
the ISR, as the spectral strength is increased in the shear range.
So
it is just the opposite situation as that discussed in section 2.2,
see also lhs of Fig.\ 1 and ref.\ \cite{amg94a}. Correspondingly,
there is also an energy pileup at the high p end of the shear range,
which may be a consequence of the constant helicity flux in this
region \cite{loh94b}. It leads to a local slope $\zeta (p) > 4/3$.

It is not our primary goal to speculate about the nature of this
crossover.
What is
important here for the discussion of apparent scaling corrections is,
that {\it positive} local scaling corrections $\delta\zeta (r) > 0$ in
r-space lead to {\it negative} $\delta\zeta (p) < 0$ in the p-space
ISR, see
Fig.\ 7b. Thus the apparent scaling corrections
$\delta\zeta^{app,p}$
have to be defined as
\begineq
\delta\zeta_2^{app, p } = \hbox{max}_{p_L \ll p \ll p_d}
 (\zeta_2 (p)) - m/3,
\label{eq53s}
\endeq
The $\rel$--dependence of $\delta\zeta^{app,p}$ is shown in Fig.8. From the
data we conclude that
\begineq
\delta\zeta^{app,p} \propto -\rel^{-3/2} < 0.
\label{eqamg9}
\endeq
For $\rel = 1500$ we have $\delta\zeta^{app,p} = -0.0082$.
This means that the qualitative
difference between the apparent scaling corrections in
$r$-- and in $p$--space is even larger for the Batchelor-- than
already for the
FMS--parametrization. For the former, even the sign of the corrections
is reversed as we go from $r$-- to $p$--space, while for the latter only
the magnitude and the asymptotic $\rel$ scaling exponent change.

Let us summarize the analysis of this section. We found that
the subtleties of
the ISR to VSR and ISR to shear range crossovers govern the
$\rel$ dependence of $\delta\zeta^{app}$. Moreover, for both
parametrizations discussed (generalized FMS and Batchelor)
the scaling corrections are quite
different in r- and in p-space. So, when analyzing the
  experimental $\rel$-dependence of $\delta\zeta^{app}$,
  one should also expect {\it different}  results in r- and in
  p-space, or, correspondingly, in the $\tau$- and $\omega$- domain.

Note, that because of the considerable lack of experimental
information about the shear range crossover, we consider eqs.\
(\ref{eq51}) and (\ref{eqamg4}) only as {\it examples}. As pointed out
above, dimensional analysis cannot distinguish between them. We think
it is worth while to experimentally or numerically study the
crossover between ISR and shear range and hope that our analyses
stimulate to do so.
This might lead to a better understanding of the $\rel$ dependence of
scaling corrections.

\section{Summary and conclusions}
\setcounter{equation}{0}
Throughout the paper we have demonstrated, that scaling properties
in $r$-- and in $p$--space can be quite different.
One could argue that in the
infinite $\rel$ limit these differences are irrelevant. This is of
course correct. Yet, as we demonstrated, for those $\rel$ which can be
achieved in experiments and even more so for the numerical ones, the finite
size corrections are considerable and it is important to know what
their influence is to be able to interpret the data correctly.
Moreover, the apparent scaling correction
$\delta\zeta^{app}$ due to shear corrections even show {\it
  asymptotically} different $\rel$ scaling behavior. For the FMS type
parametrization we had obtained
$\delta\zeta^{app,p}\propto \rel^{-3/5}$ \cite{gro94,lvo94a} and
$\delta\zeta^{app,r}\propto \rel^{-1/2}$, for Batchelor type
parametrizations
$\delta\zeta^{app,r}\propto \rel^{-3/2}$ and
$-\delta\zeta^{app,p}\propto \rel^{-3/2}$.

Comparison of the size of the scaling ranges for given $\rel$ between
experiment \cite{ans84} and our parametrizations lets us
favor a Batchelor type parametrization rather than
a parametrization of FMS type.
The latter, consisting of a power law in $p$--space with a large--$p$
exponential cutoff and a small--$p$ arctan cutoff, does not exhibit
any bottleneck energy pileups
at the ends of the p-space ISR and leads to unrealistic short
scaling ranges in the r-space structure function.
In other words,
combining all regimes discussed in this paper (VSR, ISR, shear range
and large--$r$ saturation range) we think that the $p$--space parametrization
\begineq
E(p) =
{2 E_0 \epsilon^{2/3} \over \pi} \hbox{arctan} \left(\left( {p\over p_L}
\right)^{11/3}\right)
p^{-5/3}
\left( 1 +
\left({p\over p_s}\right)^{-2/3} \right) \exp{(-p/p_d)},
\label{eq61}
\endeq
with $p_d \ge p_s \ge p_L$ is less favorable than an $r$--space parametrization
\begineq
D(r) = {\eps\over 3\nu} r^2 {r_d^{\prime 4/3}\over (r_d^{\prime 2}+r^2)^{2/3}}
             {r_s^{-2/3} \over (r_s^2+r^2)^{-1/3}}
             {L^{-4/3} \over (L^2+r^2)^{-2/3}},
\label{eq62}
\endeq
with $r_d'\le r_s \le L$,
which shows bottleneck effects in p-space.
In many isotropic turbulence experiments $r_s\approx L$ and the shear
range will be suppressed. If less isotropy is achieved in experiments,
we may have, say, $r_s\approx L/4$.
In this case, the energy pileup due to large--$r$ saturation (section 2
and ref. \ \cite{amg94a}) and the spectral strength reduction due to shear
effects will partly compensate each other at the infrared end of the
$p$--space ISR. This leads to a smaller change of the local
slope than predicted
by  eq.\ (\ref{eqn77}) or (\ref{freq5}).

Another point to be kept in mind when comparing our predictions
(\ref{eqn77}) and (\ref{freq5}) with experimental data is the issue of
averaging. While experimental scaling exponents necessarily represent data
averaged over a certain interval, we defined a {\it pointwise} (local) slope
in (\ref{eq33}) and used this basic quantity throughout the paper. In the
numerical REWA calculations \cite{gnlo94a,gnlo94b}, {\it locally averaged}
instead of pointwise slopes were determined by fitting the parametrization
(\ref{eq21}) to the data in each interval $[p/\sqrt{10}, p\sqrt{10}]$.
To estimate the effect of averaging, we performed a running average of the
pointwise slopes in fig.3 (the quantity most easily accessible in experiments)
using an averaging range of
$[\omega/\sqrt{10}, \omega\sqrt{10}]$, see fig.\ 9. As expected, the bottleneck
pileups become attenuated, so that they are now quantitatively closer to
the measured ones in fig.4.
Similar averaging algorithms could also be applied to
the other local slopes.

Finally, we mention that while eqs.
\ (\ref{eq61}) and (\ref{eq62}) explicitly distinguish between shear effects
and large $r$ saturation effects,
caused by the boundary conditions (i.e., large scale anisotropy), these effects
might be more interwoven.
In section 4.3 we had compared shear effects, finite size
effects, and the numerical REWA results. Qualitatively, they all lead
to $\delta\zeta (p) >0$ for small $p$, but the quantitative agreement was
less satisfactory. We again stress the necessity to
produce as clean shear ranges as possible in experiments. Only then, one
will be able to experimentally study the crossover phenomena associated with
the shear range, which -- together with those from VSR to ISR -- might
well be a key in understanding scaling corrections.

\vspace{1.5cm}
%\newpage
\noindent
{\bf Acknowledgements:}
We thank  A.\ Esser, G. Falkovich,
S. Grossmann, L. Kadanoff, and R.\ Kerr
for helpful suggestions,
V.\ Yakhot for supplying us with the results of
his work prior to publication, and M.\ Nelkin for bringing the history
of eq.\ (\ref{eq13}) to our attention.
D.\ L.\ heartily thanks G.\ Falkovich for
his hospitality during his stay at the Weizmann Institute of Science,
Rehovot, where part of the work was done, and
kindly acknowledges support by the Einstein foundation, by a NATO grant
through  the Deutsche Akademische Austauschdienst (DAAD),
and by DOE. A.M.--G. was supported by the Natural Sciences and
Engineering Research Council of Canada.

\vspace{3cm}
\renewcommand{\theequation}{A.\arabic{equation}}
\begin{appendix}
\section{Analytical Batchelor--type shear spectrum}
\setcounter{equation}{0}

We can analytically perform the
transformation of the Batchelor--parametrization eq.(\ref{eqamg4})
under the assumption that $r\gg r_d'$, so that
\begineq
D(r) \approx {\eps\over 3\nu} \left({r\over r_s}\right)^{2/3}
             {1 \over (r_s^2+r^2)^{-1/3}}.
\label{eqamg5}
\endeq
With this approximation we get
\begin{eqnarray}
E(p) &=& - {1\over 2\pi} \int_0^\infty pr D^{(2)}(r) \sin(pr) dr \nonumber\\
&=& {p\over 2\pi} {\eps\over 3\nu r_s^{2/3}}
{d\over dp} \left( r_s^2 - {d^2\over dp^2} \right)
 \int_0^\infty r^{2/3} (r_s^2+r^2)^{-2/3} \cos(pr)dr.
\label{eqamg6}
\end{eqnarray}
The integral in eq.(\ref{eqamg6}) can be solved \cite{pru90} to give
\begin{eqnarray}
\lefteqn{
\int_0^\infty r^{2/3} (r_s^2+r^2)^{-2/3} \cos(pr) dr  } \nonumber\\
& & = -3r_s^{1/3} {\Gamma(5/6)^2\over\Gamma(2/3)}
\phantom{}_1 F_2({5\over 6};{1\over 2},{7\over 6};{\tilde{p}^2\over 4})
+ {\sqrt{3}\over 2} \Gamma(1/3) p^{-1/3}
\phantom{}_1 F_2({2\over 3};{1\over 3},{5\over 6};{\tilde{p}^2\over 4}),
\label{eqamg7}
\end{eqnarray}
where $\tilde{p} = pr_s$. Reinserting this result into eq.(\ref{eqamg6}) we
derive after some algebra the following, rather clumsy expression,
\begin{eqnarray}
& &E(p) = {\eps r_s^{5/3} \over 3\nu} \Bigg\{
{15\over 14\pi} {\Gamma(5/6)^2\over\Gamma(2/3)} \tilde{p}^2
\bigg[ {11\over 13}
\phantom{}_1 F_2({17\over 6};{5\over 2},{19\over 6};{\tilde{p}^2\over 4}) -
\phantom{}_1 F_2({11\over 6};{3\over 2},{13\over 6};{\tilde{p}^2\over 4})
                               \nonumber\\
& & \hspace{7cm} + {187 \over 3705}
\phantom{}_1 F_2({23\over 6};{7\over 2},{25\over 6};{\tilde{p}^2\over 4})
\tilde{p}^2 \bigg] +                           \nonumber\\
& & \hspace{-.5cm}
 {\sqrt{3}\over 4\pi} \Gamma(1/3) \tilde{p}^{-7/3}
\bigg[ {28\over 27}
\phantom{}_1 F_2({2\over 3};{1\over 3},{5\over 6};{\tilde{p}^2\over 4}) -
\left( {1\over 3}
\phantom{}_1 F_2({2\over 3};{1\over 3},{5\over 6};{\tilde{p}^2\over 4}) +
{2\over 5}
\phantom{}_1 F_2({5\over 3};{4\over 3},{11\over 6};{\tilde{p}^2\over 4})
\right) \tilde{p}^2 \nonumber\\
& & \hspace{-.5cm}
+\left( {6\over 5}
\phantom{}_1 F_2({5\over 3};{4\over 3},{11\over 6};{\tilde{p}^2\over 4}) -
{9\over 11}
\phantom{}_1 F_2({8\over 3};{7\over 3},{17\over 6};{\tilde{p}^2\over 4})
\right) \tilde{p}^4 -
{108 \over 1309}
\phantom{}_1 F_2({11\over 3};{10\over 3},{23\over 6};{\tilde{p}^2\over 4})
\tilde{p}^6
\bigg]
\Bigg\}. \nonumber\\
\label{eqamg8}
\end{eqnarray}
For small $p$ the local scaling exponent
of $E(p)$ is 4/3, for large $p$ it is 2/3,
but the transition from one range to the other is non monotonous.
This is reflected in the right part of Fig.7b. In the left part of
that figure, the ultraviolet bottleneck energy pileup
can be seen in addition, which is not included in eqs.\
(\ref{eqamg5}) and (\ref{eqamg8}).

\renewcommand{\theequation}{B.\arabic{equation}}
\section{Contour integration for oscillating integrands}
\setcounter{equation}{0}

The numerical Fourier transformation
(\ref{eq27b}) of (\ref{eqamg4}) cannot straightforwardly be performed,
as the integrand is strongly oscillating and not exponentially damped.
To cope with this problem, we employ contour integration techniques
\cite{wat22}. Plugging (\ref{eqamg4}) into (\ref{eq27b}) we
obtain after some algebra
\begineq
E(p) = -{\eps \over 3\nu} r_d^{\prime 4/3} r_s^{-2/3} {p\over 4\pi} {d^3\over
  dp^3}
\int_{-\infty}^\infty {(r_s^2 +r^2)^{1/3} \exp{(ipr)}\over
(r_d^{\prime 2} + r^2)^{2/3}} dr.
\label{eqb1}
\endeq
The integral has singularities or zeros at $\pm ir_d'$ and $\pm i r_s$. Taking
the correct branch cuts and performing the corresponding contour
integration in the upper half plane, we obtain
\begineq
\int_{-\infty}^\infty {(r_s^2 +r^2)^{1/3} \exp{(ipr)}\over
(r_d^{\prime 2} + r^2)^{2/3}} dr
=
-\sqrt{3} \int_{r_d'}^\infty {|z^2 - r_s^2 |^{1/3} \exp{(-pz)}\over
(z^2- r_d^{\prime 2})^{2/3}} dz
\label{eqb2}
\endeq
or
\begineq
E(p) = {\eps \over 3\nu} r_d^{\prime 4/3} r_s^{-2/3} {\sqrt{3}p \over 4\pi}
 \int_{r_d'}^\infty { z^3 |z^2 - r_s^2 |^{1/3} \exp{(-pz)}\over
(z^2- r_d^{\prime 2})^{2/3}} dz,
\label{eqb3}
\endeq
which can now be straightforwardly integrated. Our numerical result is
displayed in Fig.7b.

\end{appendix}

\newpage

\centerline{\bf Figure Captions}

\vspace{2cm}

\noindent
Figure 1:
In the right part of the main curve we show the
energy spectrum  eq.\ (\ref{eq28}) (solid)
with, and the spectrum eq.\ (\ref{eq21})
(dashed) without the energy pileup.
In the left part the spectrum due to (\ref{eqn77}) is shown.
In the insets, the spectrum is enlarged around the energy
pileups and compared
to classical $-5/3$-scaling.

\vspace{1cm}

\noindent
Figure 2:
The local p-space scaling exponents $\zeta (p)$
%from  eq.\
%(\ref{eq10}) and eq.\ (\ref{eq16})
(solid),
and the local r-space scaling exponent $\zeta (r=1/p) $.

\vspace{1cm}

\noindent
Figure 3:
The local $\tau$- and $\omega$-space deviation from classical
scaling, $\delta\zeta (\tau = 1/\omega)$
and $\delta\zeta
(\omega )$, when assuming Batchelor kind crossovers (\ref{freq2}) and
(\ref{freq4}).

\vspace{1cm}

\noindent
Figure 4:
The {\it experimental} local $p$-space deviation from classical scaling,
$\delta\zeta (p )$, for the longitudinal energy spectrum $E_1(p)$.
This curve
corresponds to the experimental $\delta\zeta (\omega )$ via the
Taylor hypothesis.
The data are taken  from
Praskovsky and Oncley \cite{pra94} with kind permission of the
authors.

\vspace{1cm}

\noindent
Figure 5:
(a) Local p-space scaling exponents $\zeta(p)$ of the FMS and the
Batchelor type energy spectra, both with the arctan cutoff for small
$p$, see text. To allow for comparison with r-space, Fig.\ 5b, we
plotted $\zeta(p)$ versus $p_L/p$ rather than versus $p/p_L$. The
Taylor Reynolds number is $\rel=3000$, cf.\ eqs.\ (\ref{eq31},\ref{eq32}).
{}From bottom to top on the rhs of the figure, the three pairs of curves
correspond to (i) no small $p$ scaling corrections, (ii)
small $p$ scaling corrections
according to (\ref{eqamg3}) with $\delta=0.02$ and $p_b/p_L=10$, and
(iii) small $p$ scaling corrections with
$\delta=0.04$, $p_b/p_L=15$.

\noindent
(b) Local r-space scaling exponents $\zeta(r)$ for the six curves of
Fig.\ 5a. From bottom to top for both the FMS and Batchelor triple of
curves: no scaling correcions,
$\delta=0.02$, $p_b/p_L=10$ and
$\delta=0.04$, $p_b/p_L=15$.

\vspace{1cm}

\noindent
Figure 6:
Local
scaling corrections $\delta\zeta_2 (p) = \zeta_2(p) - 2/3$ due to the
REWA calculations \cite{gnlo94a,gnlo94b}, due to the infrared
bottleneck formula (\ref{eqn77}), and due to eq.\ (\ref{eq13}) (with
the arctan cutoff for small $p$) with three different values for the
unknown parameter $\alpha_2$.

\vspace{1cm}

\noindent
Figure 7:
(a) Local scaling exponents $\zeta (r) $ (solid) and $\zeta
    (p=\gamma/r)$ (dashed) when shear corrections according to
(\ref{eq51}) are present, with $Re_\lambda = 1500$ ($p_d = 3227$) and
$\alpha_2=1$. The parameter $\gamma$ serves to shift the $p$--space curve
slightly to ensure that the minima of the two curves coincide.

\noindent
(b) As in (a), but now shear corrections according to (\ref{eqamg4}) and
    its Fourier transform. Again, we chose $Re_\lambda = 1500$,
($p_d^\prime =
    r_d^{ -1}
    = 676$).

\vspace{1cm}

\noindent
Figure 8:
Double logarithmic plot of
$\delta\zeta^{app,r} (Re_\lambda)$ and
$\delta\zeta^{app,p} (Re_\lambda)$ for FMS ($\alpha_2=1$) and
Batchelor parametrization.

\vspace{1cm}

\noindent
Figure 9:
Same as in fig.\ 3, but now in addition the averaged local slopes. The
averaging range is $[\omega/\sqrt{10}, \omega\sqrt{10}]$.

\newpage

\noindent
$^*$ On leave of absence from Fachbereich Physik, Universit\"at
Marburg, Renthof 6, D-35032 Marburg.

%\bibliography{literatur}

\end{document}